\documentclass{icrc2009}

\usepackage{graphicx}   
\usepackage{caption}    
\usepackage[font=footnotesize]{subfig} 
\usepackage{fixltx2e}
\usepackage{url}

\newcommand{\shorttitle}[1]%
{\markboth{Proceedings of the 31\MakeLowercase{$^{st}$} ICRC, {\L}\'{o}d\'{z} 2009}{#1} }
\newcommand{\etal}{\MakeLowercase{\textit{et al. }}} 

\begin{document}
\title{Underwater acoustic detection of UHE neutrinos with the ANTARES experiment}

\author{\IEEEauthorblockN{Francesco Simeone, on behalf of the ANTARES Collaboration.\IEEEauthorrefmark{1}}
                            \\
\IEEEauthorblockA{\IEEEauthorrefmark{1}University "La Sapienza" and INFN Sez. Roma.}}

\shorttitle{Simeone \etal Acoustic neutrino detection}
\maketitle

\begin{abstract}
 The ANTARES Neutrino Telescope is a water Cherenkov detector composed of an array of approximately 900 
 photomultiplier tubes in 12 vertical strings, spread over an area of about 0.1 km$^{2}$ with an 
 instrumented height of about 350 metres. ANTARES, built in the Mediterranean Sea, is the biggest 
 neutrino telescope operating in the northern hemisphere. Acoustic sensors (AMADEUS project) have been 
 integrated into the infrastructure of ANTARES, grouped in small arrays, to evaluate the feasibility 
 of a future acoustic neutrino telescope in the deep sea operating in the ultra-high energy regime.
 In this contribution, the basic principles of acoustic neutrino detection will be presented. 
 The AMADEUS array of acoustic sensors will be described and the latest results of the project 
 summarized.\\
\end{abstract}

 
\section{Introduction}
 Almost everything we know about the Universe came from its observation by means of electromagnetic radiation. 
 Using photons as observation probe it has been possible to discover very energetic sources. 
 However, photons are highly absorbed by matter and so their observation only allows us to directly 
 obtain information of the surface process at the source. Moreover energetic photons interact with 
 the infrared photon background and are attenuated during their travel from the source toward us.\\
 Observation of the proton component of cosmic rays can give information about the sources but, 
 since they are charged, low energy protons are deflected by the magnetic galactic fields and loose 
 the directional information that would allow us to point back to their source. Ultra high energy protons 
 are slightly deflected by magnetic fields and in principle could be a good probe for the high energy 
 Universe. Unfortunately, as pointed out by Greisen-Zatsepin-Kuz'min \cite{GZK1} \cite{GZK2}, the 
 proton interactions with the CMBR(GZK effect) will reduce the proton energy and make them not useful as 
 astroparticle probes for the high energy Universe.\\
 In order to directly observe the physical mechanism of distant and energetic sources we need 
 to use a neutral, stable and weakly interacting messenger: the neutrino. The interest in studying 
 such high energy sources arises from the fact that much of the classical astronomy is related to 
 the study of the thermal radiation, emitted by stars or dust, while the non thermal energy density 
 in the Universe is roughly equal to the thermal one and it is assumed to play a relevant role in its 
 evolution.\\  
 The experimental techniques proposed to identify the cosmic neutrino signatures are mainly three: 
 the detection of Cherenkov light originating from charged leptons produced by neutrino interactions 
 in water or ice; the detection of acoustic waves produced by neutrino induced energy deposition in water, 
 ice or salt; the detection of radio pulses following a neutrino interaction in ice or salt.\\

 \begin{figure*}[ht]
  \centering
  \includegraphics[width=0.7\textwidth]{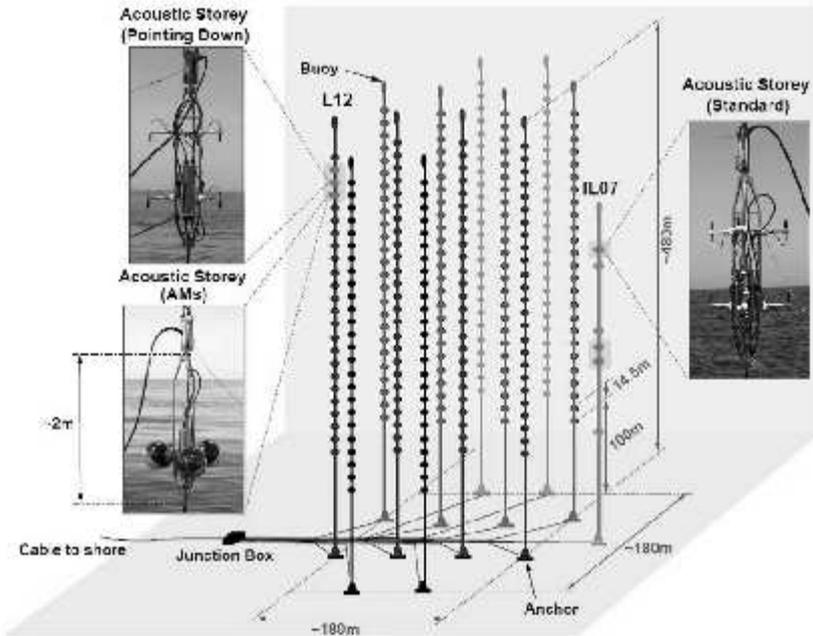}
  \caption{Acoustic storeys are present in four different configurations; two of
 them contain commercial hydrophones (one upward and one downward
 looking), one contains custom built hydrophones and the last is made of
 sensors inside three glass spheres. }
  \label{ANT_LAY}
 \end{figure*}
 
\section{Acoustic neutrino detection}
 The acoustic detection technique of neutrino-induced cascades, in water or ice, is based on the 
 thermo-acoustic effect \cite{TERMO1} \cite{TERMO2}. The cascade energy is deposited in a narrow region of the medium, 
 inducing a local heating and resulting in a rapid expansion of the water (or ice). The simultaneous 
 expansion of the medium along the shower leads to a coherent sound emission in the plane 
 perpendicular to the shower axis.\\
 Simulations performed by many authors \cite{SIMUL1} \cite{SIMUL2} show that the bipolar acoustic neutrino pulse is tens 
 of microseconds long and has a peak-to-peak amplitude of about 100~mPa at 1km distance if originated 
 by a shower of 10$^{20}$~eV. The signal is largely collimated in the plane perpendicular to the 
 shower axis and its amplitude decreases by almost two order of magnitude in few degrees.\\
 The acoustic detectors are composed of many acoustic sensors distributed in a wide instrumented 
 volume; by measuring the acoustic induced neutrino pulse with several sensors it will be possible 
 to infer the shower direction. The interest in this technique is related to the high attenuation 
 length ($\sim$km) of the sound in water (or ice). Consequently it is possible to instrument a large volume 
 using a relatively low number of sensors.\\

\section{AMADEUS project}
 The AMADEUS \cite{AMADEUS} (Antares Modules for Acoustic DEtection Under the Sea) project is fully integrated into 
 the ANTARES \cite{ANTARES} \cite{ANTARES1} \cite{ANTARES2} Cherenkov neutrino telescope (Fig.~\ref{ANT_LAY});
 Its main goal is to evaluate the feasibility 
 of a future acoustic neutrino telescope in the deep sea operating in the ultra-high energy regime.\\
 ANTARES is located in the Mediterranean Sea about 40~km south of Toulon (France) at a depth of about 
 2500m. It is composed of 12 vertical structures, called lines and labeled L1 - L12. An additional 
 line (IL07) is equipped with several instruments used for environmental monitoring. Each detection 
 line holds up to 25 storeys, each of them contains three photomultipliers (PMTs) and the electronics needed to 
 acquire the PMT signals and send them to shore. The storeys are vertically separated by 14.5~m 
 starting at a height of about 100m above sea floor. Each line is fixed to the sea floor by an 
 anchor and held vertically by a buoy.\\ 
 Three special storeys (Acoustic Storeys) are present on both L12 and IL7, each storey contains six 
 acoustic sensors and the electronics used to acquire, pre-process and send the samples to the onshore laboratory. The 
 hydrophones signals are amplified up to a sensitivity of 0.05V/Pa, filtered using a bandpass filter 
 from 1~kHz to 100~kHz and sampled at 250~ksps with a resolution of 16~bit. The acquisition system is 
 able to produce up to 1.5~TB per day; in order to select interesting signals and reduce the data 
 rate as well as the storage requirements, three online triggers are implemented at level of the 
 acoustic storey:
 \begin{itemize}
  \item Minimum bias filter, that stores 10~s of samples every hour. 
  \item Threshold filter, that store signals of amplitude greater than the selected threshold. 
  \item Matched filter, that store signals of cross correlation with the expected bipolar signal 
        greater than the selected threshold.
 \end{itemize}
 These filters reduce the data sent to the onshore laboratory by two orders of magnitude. The AMADEUS project is fully
 integrated into the ANTARES data acquisition system, in particular all the samples acquired by AMADEUS are tagged with
 an absolute time, common to the whole experiment, with a precision better than 1ns. This allows to correlate acoustic
 signals acquired in different parts of the apparatus and represents, at the moment, an unique underwater hydrophone
 array acquiring synchronously.\\
 \begin{figure*}[ht]
  \setcounter{figure}{2}
  \centering
  \includegraphics[width=0.75\textwidth]{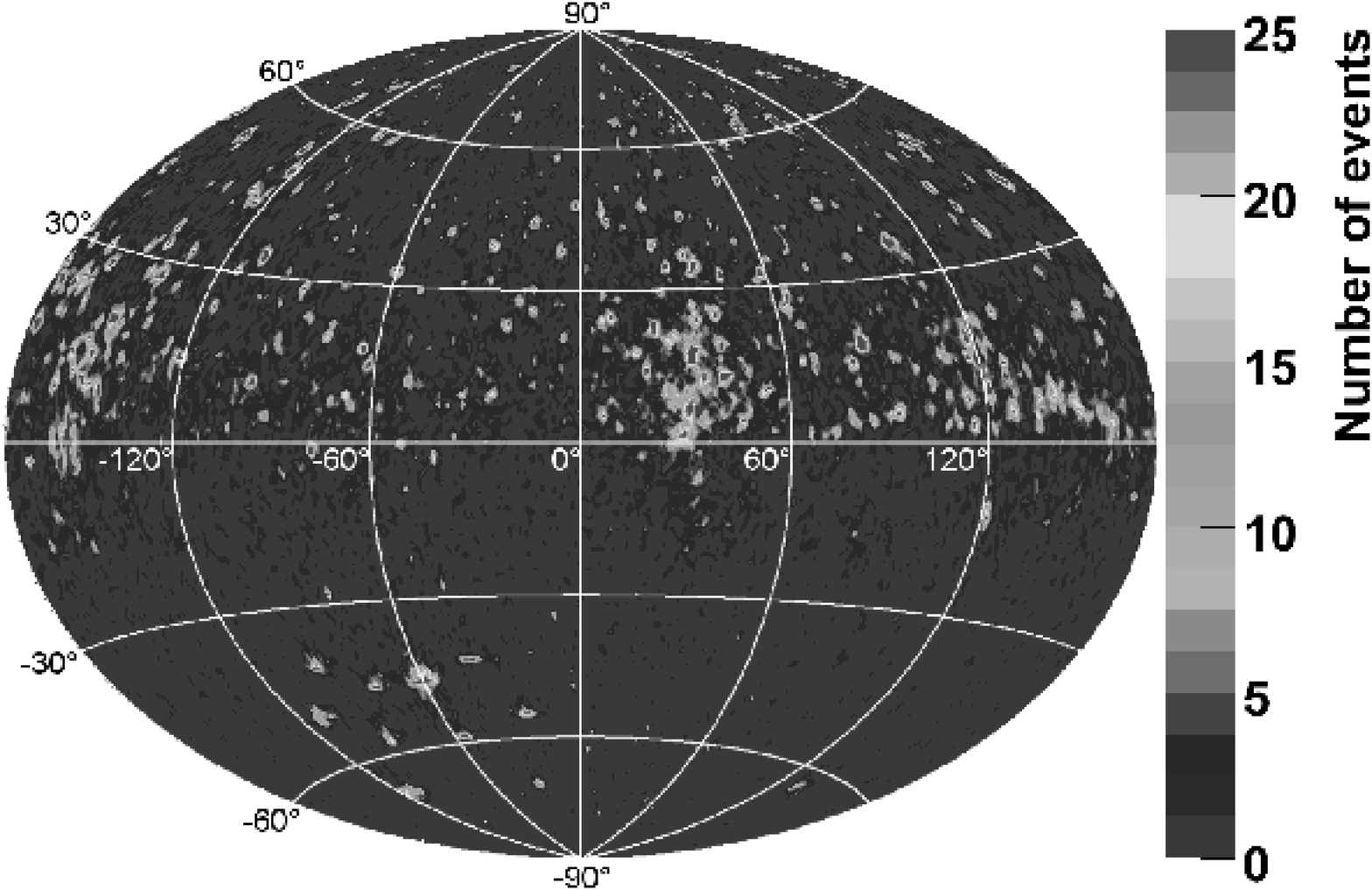}
  \caption{Mapping of the arrival directions of transient signals. The map is produced using the data acquired by the
 second storey on IL07 about 180m above the sea-bed. The origin is defined by the westward direction with respect to
 the horizontal of that storey.
 90$^{\circ}$ (-90$^{\circ}$) in longitude corresponds to north (south), 90$^{\circ}$ (-90$^{\circ}$) in
 latitude to vertically downward-going acoustic signals (upward-going). }
  \label{MAP_DIR}
 \end{figure*}

\section{First AMADEUS results}
 The underwater enviroment is an highly noisly ambient: thus the characterization of this background is of
 fundamental importance to develop detection algorithms able to identify neutrino induced acoustic signals.
 One of the main opportunity given by the AMADEUS system, is to study this background.\\
 The noise sources can be classified in two main group: 
\begin{itemize}
  \item Transient signals (like the ones expected by the UHE neutrinos). These signals contains a finite amount of energy and have a finite duration in time.
  \item Stationary random signals. These signals have statistical properties that are invariant with respect to a
 translation in time and are often characterized by their power spectral density (PSD).
 \end{itemize} 
 In the following subsections the contribution of these two kinds of signals will be discussed.

 \subsection{Transient Signals}
 Each storey of the AMADEUS project is an array of hydrophones; taking advantage of these geometry, 
 it is possible to reconstruct the arrival direction of the pressure wave on the storey using a 
 technique called beam-forming.\\
 The Beam-forming is a MISO (Multi Input Single Output) technique \cite{BEAM} originally developed to 
 passively detect submarines with an underwater array of hydrophones. This technique analyses the 
 data acquired by an array of sensors to compute the arrival direction of the incident waves. The 
 array of sensors samples the wave-front in space and time and the samples of different sensors are 
 combined, using the proper time delays, to sum up coherently all the waves arriving on the array 
 from a specific direction; this will lead to an increase of the signal to noise ratio (SNR) since the
 spatially white noise 
 will be averaged out. Since we don't know the arrival direction of the incident wave we need to 
 compute the time delays for many different arrival directions and produce an angular map of the 
 incident pressure wave as the one  reported in figure~\ref{MAP_ANG}. The aliases that are present in this map 
 are due to delays (directions) for which two or more of the six hydrophones sum coherently. The only 
 direction for which all the hydrophones sum up coherently is the true one.\\
\begin{figure}[htb]
  \setcounter{figure}{1}
  \centering
  \includegraphics[width=0.4\textwidth]{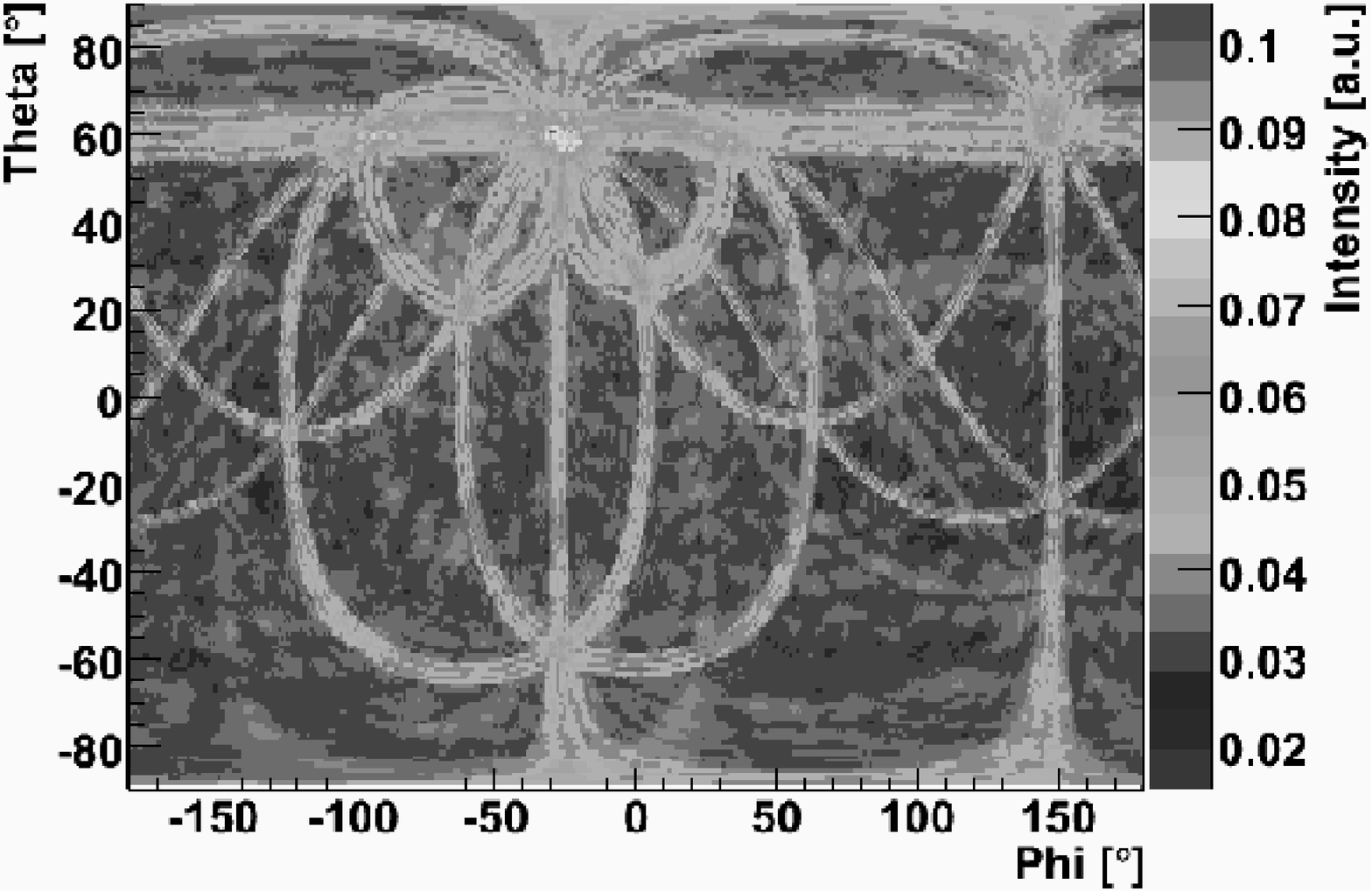}
  \caption{Example of beamforming output used to reconstruct a transient signal direction. The beamforming output
 is a function of the number of hydrophones whose signals sum up coherently, assuming that the transient signals
arrive from that specific direction.}
  \label{MAP_ANG}
 \end{figure}
 Using the directional information provided by three or more arrays it is possible to reconstruct the 
 acoustic source position. The directional information can be used in conjunction with ray-tracing 
 techniques that allows us to easily take into account the sound propagation in water.\\ 
 In figure~\ref{MAP_DIR} qualitative mapping of the arrival directions of transient acoustic signals are shown.
 The data sample has been collected with the minimum bias filter during a time period of about 6 
 months; the figure shows the directions of all reconstructed signals which have an amplitude greater 
 than eight times the standard deviation of the ambient noise. The majority of the reconstructed 
 acoustic signals are received from directions in the upper hemisphere; this is consistent with the expectations 
 since the major sources of transient noise are due to biological and anthropological activities. 
 The few souces visible in the lower left part of the sphere resemble the layout of the ANTARES strings.
 Those sources are the pingers of the ANTARES acoustic positioning system,
 emitting acoustic signals at the bottom of each line.\\
 \subsection{Correlation between underwater noise and weather condition.}
 An analysis correlating the ambient acoustic noise level, measured with AMADEUS and the surface weather data was
 performed for the whole year 2008. The data used in this analysis are the minimum bias ones and the weather condition
 were measured by 5 station around the ANTARES site:
 \begin{itemize}
  \item Station 1: Toulon.
  \item Station 2: Cap Cepet.
  \item Station 3: Hy$\grave{e}$res.
  \item Station 4: Porquerolles.
  \item Station 5: Toulon/Ile du Levant.
 \end{itemize}
 For each station a daily average of the wind speed and other weather observables are available; in this analysis only the
 mean wind speed was used. The noise level for a data sample acquired by AMADEUS is evaluated by integrating over the
 frequency range 1-50~kHz, the mean PSD is calculated using samples of 8.4 seconds duration.\\
\begin{figure}[htb]
  \setcounter{figure}{3}
  \centering
  \includegraphics[width=0.41\textwidth]{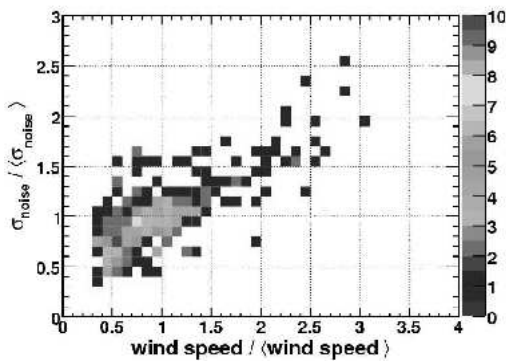}
  \caption{The relative variation of the wind speed, measured by the weather station 2, and the noise level measured by
 the sensor 17 (IL07). This weather station is located near the sea so its wind measurements are well suited for this
 kind of analysis.}
  \label{CORR_2}
 \end{figure}
 For better comparability between wind speed and noise level, the daily variation with respect to the annual mean is used.
 The correlation seems to became more evident for high wind speed (see figure~\ref{CORR_2}).\\
\begin{figure}[htb]
  \centering
  \includegraphics[width=0.4\textwidth]{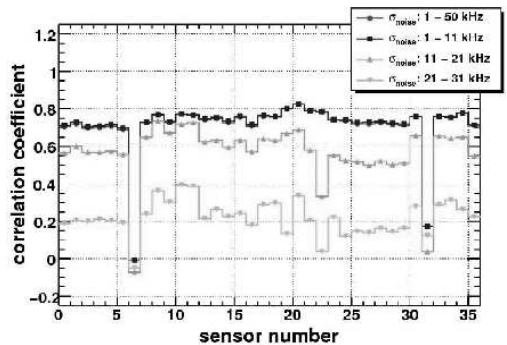}
  \caption{Correlation coefficient evaluated using different frequency bands. Sensors 6 and 31 have been defective from
 the beginning of the data taking and thus show no correlation with the weather conditions. Sensor 22 has a non-ideal
 coupling with the glass sphere and is not sensitive to the higher frequency range.}
  \label{CORR_3}
 \end{figure}
 For frequencies below 21~kHz a strong correlation ($\geq$50\%) is found; for higher frequency the ambient noise decreases
 and the system noise starts to becomes significant as a consequence the correlation with the weather condition
 decreases (see figure~\ref{CORR_3}).
 No systematic different observation is found using the other weather station data; the significance of the correlation
 is lower as the stations are situated in-land and typically measure lower wind speed.\\

\section{Conclusions} 
 The AMADEUS project is fully operating and the data analysis is going on; this system offers a unique opportunity
 to study the underwater noise in order to develop and test acoustic triggers, able to discriminate the neutrino-induced
 signals from underwater background.\\
 The matched filter is the optimum linear filter to discriminate a signal of known shape from a 
 stochastic background of known spectrum. The beam-forming technique allows us to further increase 
 the SNR by a factor that is equal to the number of hydrophones used in the phased array. Moreover 
 the beam-forming technique allows to reconstruct the arrival direction of the incident pressure wave. 
 The directional information, combined with a ray-tracing technique, permits an easy reconstruction 
 of the interaction point. The combination of these techniques could extend the energy range of 
 acoustic detectors and may increase the possibility to measure combined events in an hybrid 
 (optic-acoustic) detector.\\

\end{document}